\documentclass[a4paper,fleqn,usenatbib]{mnras}
\usepackage{graphicx}
\usepackage{amssymb}
\usepackage{amsmath}
\usepackage{ae,aecompl}
\usepackage{natbib,times}
\usepackage{lscape}

\newcommand{\mc}{\multicolumn}

\begin{document}
\title[Massive galaxy clusters at high redshifts]
      {A sample of 1959 massive galaxy clusters at high redshifts}

\author[Wen \& Han]
{Z. L. Wen$^{1,2}$\thanks{E-mail: zhonglue@nao.cas.cn}
  and J. L. Han$^{1,2,3}$ \thanks{E-mail: hjl@nao.cas.cn}
\\
1. National Astronomical Observatories, Chinese Academy of Sciences, 
20A Datun Road, Chaoyang District, Beijing 100012, China\\
2. CAS Key Laboratory of FAST, NAOC, Chinese Academy of Sciences,
           Beijing 100101, China \\
3. School of Astronomy, University of Chinese Academy of Sciences,
           Beijing 100049, China 
}

\date{Accepted XXX. Received YYY; in original form ZZZ}

\label{firstpage}
\pagerange{\pageref{firstpage}--\pageref{lastpage}}
\maketitle


\begin{abstract}
We identify a sample of 1959 massive clusters of galaxies in the
redshift range of $0.7<z<1.0$ from the survey data of Sloan Digital
Sky Survey (SDSS) and Wide-field Infrared Survey Explorer
(WISE). These clusters are recognized as the overdensity regions
around the SDSS luminous red galaxies, having a richness greater than
15 or an equivalent mass $M_{500} \ge 2.5\times10^{14}M_{\odot}$.
Among them, 1505 clusters are identified for the first time, which
significantly enlarge the number of high-redshift clusters of
$z>0.75$. By comparing them with clusters at lower redshifts, we
confirm that richer clusters host more luminous brightest cluster
galaxies (BCGs) also at high redshifts, and that the fraction of blue
galaxies is larger in clusters at higher redshifts. A small fraction
of BCGs show ongoing star formation or active nuclei. The number
density profile of member galaxies in stacked samples of clusters 
shows no significant redshift evolution.
\end{abstract}


\section{Introduction}

The formation and evolution of clusters have become a hot subject in
astrophysics for decades \citep[see a review in][]{kb12}.  As the
largest gravitationally bound system, clusters of galaxies are formed
at knots of the cosmic web. Clusters at higher
redshifts possess more star-forming galaxies with a blue color, which
is known as the Butcher--Oemler effect \citep{bo78,bo84,doc+97}. 
The member galaxies of clusters in the
local universe are dominated by red early-type galaxies
\citep{roo69,bts87,fs91}. However, the progenitors of clusters known
as protoclusters are usually dominated by star-forming galaxies at
redshifts $z>2$ \citep{mot+04,vrm+07,crs+11}. The
stellar population in most brightest cluster galaxies (BCGs) was also
formed at redshifts $z>2$ and evolved passively afterwards
\citep{wad+08,ses+08,wh11}, but stellar mass of BCGs increases
significantly with time by accreting satellite galaxies
\citep{db07,lsm+12,blm+16}. Evolution studies of cluster properties
require a large sample of clusters covering a wide redshift range.

Recent years, the Sloan Digital Sky Survey \citep[SDSS,][]{yaa+00}
provides a great data base to identify large samples of galaxy
clusters to a redshift of $z\sim0.8$ \citep{kma+07b,whl09,hmk+10,
  spd+11,whl12, rrb+14,ogu14,wh15,rrh+16,bsp+18}. Massive clusters up
to redshifts of $z\sim1.6$ have been identified and
spectroscopically confirmed by optical and infrared observations
\citep{seb+05,cst+07,kcz+09,wmy+09,pmw+10,ccm+10,sbg+12}. Larger
samples of high-redshift clusters have been identified from deep
multi-band photometric surveys. \citet{gy05} identified 429 cluster candidates at
redshifts $0.2 <z <1.4$ from the 10.13 square degree area of
Red-Sequence Cluster Survey. \citet{ghi+08} identified 16 galaxy
clusters at redshifts $0.9 < z < 1.7$ from the 0.4 square degree AKARI
data. \citet{zrw+07} identified 12 cluster candidates at redshifts
$1.23\le z \le 1.55$ from the 0.66 square degree area of Cosmic
Evolution Survey (COSMOS). \citet{vcb+06} identified 13 clusters at
redshifts $0.61\le z\le1.39$ from the 0.5 square degree area of UKIRT
Infrared Deep Sky Survey Early Data Release. \citet{ebg+08} identified
335 galaxy cluster and group candidates using the 4.5 $\mu$m-selected
sample of objects from the 7.25 square degree area of Spitzer Infrared
Array Camera Shallow Survey. \citet{wh11} identified 631 clusters, 202
clusters, 187 clusters and 737 clusters in a redshift range of
$0.1<z<1.6$ respectively from the 35 square degree area of
Canada-France-Hawaii Telescope (CFHT) Wide survey, the 3.2 square
degree area of CFHT Deep survey, the 2.0 square degree area of COSMOS,
and the 33 square degree area of Spitzer Wide-area InfraRed
Extragalactic survey.  \citet{oll+18} identified 1921 clusters at
redshifts $0.1<z<1.1$ from the 232 square degree area of Hyper
Suprime-Cam (HSC) Subaru Strategic Program. Searching for these
high-redshift clusters from optical and infrared deep surveys were
made over a sky area of only a few hundred square degrees, and
only a few hundred massive clusters with a mass $>10^{14}~M_{\odot}$
at redshifts $z>0.75$ have been recognized.
In addition, a few tens X-ray massive clusters at redshifts $z>0.75$
were found from many small fields of deep X-ray surveys
\citep{jse+98,pap+11,fbn+11,ssb+12,cal+14,fwt+10,pcg+16}, and another
a few tens clusters of redshifts $z>0.75$ were found via the
Sunyaev-Zel'dovich (SZ) effect on the sky surveys of cosmic microwave
background by Planck satellite, Atacama Cosmology Telescope (ACT), and
South Pole Telescope (SPT) \citep{maa+11,hhm+13,bsd+15,plancksz16}.

Wide-field Infrared Survey Explorer \citep[WISE,][]{wem+10} is an all
sky survey in four mid-infrared bands. The currently largest sample of
47,600 clusters with a redshift of $0.025<z<0.4$ have been identified
from the combined data of Two Micron All Sky Survey, WISE, and
SuperCOSMOS covering 28,000 square degrees of the sky \citep{why18}.
Because of the 1.6 $\mu$m bump in the spectral energy distributions,
the magnitudes of early-type galaxies in the WISE bands vary with
redshift significantly slower than optical magnitudes in the redshift
range of $0.5<z<1.5$ \citep{ydt+13}. Hence, the WISE data have a good
advantage to detect a large sample of galaxy clusters up to a redshift
of $z\sim1$. By using WISE and SDSS photometric data and follow-up
spectroscopy, \citet{ggs+12} made the Massive Distant Clusters of Wise
Survey, and reported their first discovery of a massive high-redshift
cluster at a redshift of $z=0.99$. Another 20 massive clusters at
redshifts of $0.7<z<1.2$ have been confirmed later
\citep{sgb+14,gdb+15}.

In this paper, we present a large sample of 1959 massive clusters of
galaxies of $0.7<z<1$ identified from the data of SDSS and WISE
covering $\sim$10000 square degrees of the sky, and 1505 of them are
recognized for the first time.
In Section 2, we first describe the galaxy data and the identification
procedures for high-redshift galaxy clusters. In Section 3, we study
the evolution of cluster properties, including color evolution of
BCGs, the fraction of blue galaxies in clusters and the number density
profile of cluster member galaxies. A short summary is presented in
Section 4.

Throughout this paper, we assume a flat Lambda cold dark matter
cosmology taking $H_0=70$ km~s$^{-1}$ Mpc$^{-1}$, $\Omega_m=0.3$ and
$\Omega_{\Lambda}=0.7$.

\section{Clusters of galaxies identified from the SDSS and WISE}

The high-redshift massive clusters are identified from the WISE data
and the SDSS data by following a few steps. The luminous red galaxies
(LRGs) with spectroscopic redshifts in the SDSS are first taken as BCG
candidates. The photometric data of the SDSS are used to discard stars
and low-redshift galaxies in the WISE catalogue. Galaxy clusters are
then recognized as the overdensity regions in the remaining WISE
source catalogue around the LRGs.

\subsection{Galaxy data}

The SDSS\footnote{http://www.sdss.org/} performs the optical photometric
survey and the follow-up spectroscopic survey in the northern sky
\citep{yaa+00}. The photometric data have been taken at five broad
bands ($u$, $g$, $r$, $i$, and $z$) covering 14,000 square degrees of
the sky \citep{dr8+11} reaching a limit of $r=22.2$ mag
\citep{slb+02}. Among 278 million sources from the photometric data,
112 million sources have been classified as stars, and other 166
million sources are classified as galaxies. All galaxies have 
their photometric redshifts estimated already \citep{bcl+14}.
The SDSS spectroscopic survey has been made to the galaxies with an
extinction-corrected Petrosian magnitude of $r<17.77$ for the main
galaxy sample \citep{swl+02} and the LRGs of $r<19.5$ selected by using
the $gri$ magnitudes in the SDSS-I and II phases \citep{eag+01}. The
SDSS-III Baryon Oscillation Spectroscopic Survey (BOSS) has completed
the spectroscopic observations of the LRGs of $i<19.9$ to a redshift
of $z\sim0.7$ covering $\sim10000$ square degrees of the sky
\citep{dr8+11}. The SDSS-IV extended BOSS (eBOSS) further observed the
LRGs in the redshift range of $0.6<z<1.0$ selected from the
photometric data of the SDSS and WISE in the BOSS region covering 7500
square degrees of the sky \citep{pln+16}.
In this work, we take the spectroscopic redshifts of galaxies from the
latest SDSS Data Release 14 \citep[DR14,][]{dr14+17}, include the data
of SDSS-III covering 10000 square degrees of the sky and the latest
eBOSS data covering 2480 square degrees of the sky. According our
previous result \citep{whl12}, most of BCGs are LRGs. We therefore
take the 296,832 LRGs as BCG candidates, which all have spectroscopic
redshifts observed ready. Contaminating active galactic nuclei are
removed by setting the `spectra class = GALAXY' and `zWaring = 0'.

The WISE survey\footnote{http://irsa.ipac.caltech.edu/Missions/wise.html}
observes the whole sky in four mid-infrared bands \citep{wem+10}: $W1$
(3.4 $\mu$m), $W2$ (4.6 $\mu$m), $W3$ (12 $\mu$m), and $W4$ (22 $\mu$m)
with 5$\sigma$ magnitude limits of 17.1, 15.7, 11.5, and 7.7 mag in the
Vega system for point sources, respectively.
In the SDSS BOSS region, we get 84 million WISE sources with a
detection signal-to-noise greater than 10 in the $W1$ band, the
fraction of saturation pixels of `w1sat $< 5\%$', and the
contamination flag of `ccflag = 0000'. This magnitude limit in the
$W1$ band corresponds to $0.7\,L^{\ast}$ at $z=0.7$ and slightly
increases to $0.9\,L^{\ast}$ at $z=1$, here $L^{\ast}$ is the
characteristic luminosity of galaxies in clusters \citep{mgb+10}.
We convert the Vega magnitudes to AB magnitudes for the WISE sources
by adding the offsets of 2.699, 3.339, 5.174, and 6.620 for $W1$,
$W2$, $W3$ and $W4$ magnitudes, respectively \citep{jcm+11}. 

For high-redshift galaxies, the $W1$-band photometric data of WISE are
deeper than the SDSS data due to the slow change of galaxy magnitude
with redshift \citep{ydt+13}. We therefore identify high-redshift
clusters from the WISE source catalogue, after foreground objects are
discarded by using the SDSS photometric data. Considering the
resolution of WISE data, we first remove the WISE sources within 3
arcsec of the known SDSS stars, and then remove the WISE sources
within 3 arcsec of the known SDSS galaxies with a photometric redshift
of $z<0.65$ (otherwise $z<0.7$ but considering the uncertainties
photometric redshifts).
After these foreground objects removed, we finally get only 27 million
sources from the foreground-cleaned SDSS-WISE data, which are used for
further identification of high-redshift massive clusters of $0.7<z<1$
(see Section~\ref{algorithm}).

\begin{table*}
\begin{minipage}{160mm}
\caption[]{The 1959 massive clusters of galaxies identified from the SDSS and WISE.}
\begin{center}
\setlength{\tabcolsep}{1mm}
\begin{tabular}{rrrcccccccccc}
\hline
\mc{1}{c}{Name}&\mc{1}{c}{R.A.} & \mc{1}{c}{Dec.} & \mc{1}{c}{$z$} & 
\mc{1}{c}{$i_{\rm BCG}$} & \mc{1}{c}{$W1_{\rm BCG}$} &  \mc{1}{c}{SNR} & \mc{1}{c}{$r_{500}$} &
\mc{1}{c}{$L_{500}$} & \mc{1}{c}{$R_{L*,500}$} & \mc{1}{c}{$M_{500}$} &\mc{1}{c}{$N_{\rm sp}$} & 
\mc{1}{c}{Other catalogues} \\
\mc{1}{c}{(1)} & \mc{1}{c}{(2)} & \mc{1}{c}{(3)} & \mc{1}{c}{(4)} & \mc{1}{c}{(5)} & 
\mc{1}{c}{(6)} & \mc{1}{c}{(7)} & \mc{1}{c}{(8)} & \mc{1}{c}{(9)} & \mc{1}{c}{(10)} &
\mc{1}{c}{(11)} & \mc{1}{c}{(12)} & \mc{1}{c}{(13)} \\
\hline
WH J000133.5$-$085212 & 0.38965 & $-8.86999$ & 0.7938 & 19.51 & 17.25 & 8.12 & 0.75 & 13.61 & 15.73 & 2.59 & 10 &     \\
   J000223.3$-$051523 & 0.59721 & $-5.25636$ & 0.7044 & 19.52 & 17.55 & 5.78 & 0.73 & 15.36 & 15.47 & 2.54 & 10 & WH15\\
WH J000309.8$+$082208 & 0.79096 & $ 8.36882$ & 0.7370 & 19.61 & 18.02 & 6.42 & 0.76 & 17.36 & 18.40 & 3.06 & 12 &     \\
WH J000311.3$+$020653 & 0.79696 & $ 2.11459$ & 0.8733 & 20.48 & 18.20 & 7.05 & 0.57 & 13.36 & 17.35 & 2.87 & 10 &     \\
WH J000354.1$+$304223 & 0.97556 & $30.70643$ & 0.7398 & 19.86 & 17.62 & 6.46 & 0.86 & 18.50 & 19.69 & 3.29 & 18 &     \\
WH J000410.1$+$061513 & 1.04228 & $ 6.25357$ & 0.8071 & 19.95 & 17.87 & 7.09 & 0.68 & 15.02 & 17.71 & 2.94 & 12 &     \\
WH J000419.7$+$335417 & 1.08207 & $33.90479$ & 0.7258 & 19.47 & 17.70 & 7.26 & 0.76 & 17.18 & 17.90 & 2.97 & 14 &     \\
WH J000452.1$+$261003 & 1.21715 & $26.16741$ & 0.8176 & 20.22 & 17.90 & 5.54 & 0.80 & 18.56 & 22.22 & 3.75 & 16 &     \\
WH J000546.7$+$022256 & 1.44453 & $ 2.38212$ & 0.8397 & 20.52 & 17.45 & 7.98 & 0.77 & 13.84 & 17.12 & 2.83 &  9 &     \\
WH J000800.6$+$095535 & 2.00246 & $ 9.92629$ & 0.7236 & 19.72 & 17.74 & 5.40 & 0.86 & 21.94 & 22.78 & 3.86 & 17 &     \\
\hline
\end{tabular}
\end{center}
{Note.
Column 1: Cluster name with J2000 coordinates of cluster and the suffix of ``WH'' indicating the new identification in this paper; 
Column 2 and 3: Right Ascension (R.A. J2000) and Declination (Dec. J2000) of cluster BCG (in degree);
Column 4: spectroscopic redshift of the cluster; 
Column 5--6: BCG magnitudes in $i$ and $W1$, respectively;
Column 7: signal-to-noise ratio for cluster detection;
Column 8: cluster radius, $r_{500}$, in Mpc; 
Column 9: $L_{500}$ in units of $L^{\ast}$;
Column 10: cluster richness.
Column 11: $M_{500}$ in units of $10^{14}~M_{\odot}$;
Column 12: number of member galaxy candidates within $r_{500}$; 
Column 13: other catalogue containing the cluster: RCS \citep{gy05}, 
SDSSCGB \citep{mpe+09}, MCXC \citep{pap+11}, GMB+11 \citep{gmb11},
WH11 \citep{wh11}, DAC+11 \citep{dac+11}, WHL12 \citep{whl12}, 
 XMMXCS \citep{mrh+12}, SKT+12 \citep{skt+12}, ACT \citep{hhm+13,hhs+18},
 2XMMi-SDSS \citep{tsl13,tsl14}, WH15 \citep{wh15},
BSG+15 \citep{bsg+15}, DAB+15 \citep{dab+15}, SWXCS \citep{ltt+15}
CAMIRA-HSC \citep{oll+18}.\\
(This table is available in its entirety in a machine-readable form.)
}
\label{tab1}
\end{minipage}
\end{table*}

\subsection{The scaling relations for cluster radius and mass}
\label{scale}

The radius and mass are two fundamental parameters for galaxy
clusters, and their scaling relations to observational data are the
base for cluster identification. The cluster radius, $r_{500}$, is the
radius within which the mean density of a cluster is 500 times of the
critical density of the universe, and $M_{500}$ is the cluster mass
within $r_{500}$. Before identifying clusters, we first have to find
out the scaling relations for estimating cluster radius and mass by
using the cleaned SDSS-WISE data (without galaxies of $z<0.65$).

As done in our previous papers \citep{whl12,wh15}, we get the scaling
relations between the total $W1$-band luminosities and known $r_{500}$
(and $M_{500}$) of clusters compiled from literature. As listed in
Table~\ref{app}, we compile the cluster masses $M_{500}$ for 45 known
clusters at redshifts $0.7\le z \lesssim 1$. These clusters are
located in the SDSS region and have the SDSS-WISE data as discussed
above. We recognize the member galaxy candidates in the cleaned
SDSS-WISE data for each galaxy cluster, supposing that they are
fainter than the BCG. The total $W1$-band luminosity is calculated by
summing the luminosities of all member galaxy candidates. Follow
\citet{whl12} and \citet{wh15}, the local background is estimated from
an annulus of projected distance between 2--4 Mpc from the BCG
candidate, and then subtracted from the total $W1$-band luminosity. As
shown in the appendix, we find that the radius $r_{500}$ is related to
total $W1$-band luminosity within a radius of 1 Mpc from a BCG $L_{\rm
  1Mpc}$ (in units of $L^{\ast}$) and redshift $z$ by
\begin{eqnarray}
\log r_{500}&=&0.45\log L_{\rm 1Mpc}-(0.81\pm0.12)\nonumber\\
&&+(0.57\pm0.29)\log (1+z). 
\label{r500z}
\end{eqnarray}
We then calculate the total $W1$-band luminosity $L_{500}$ (also in
units of $L^{\ast}$) within the radius of $r_{500}$, and then find
that cluster mass $M_{500}$ (in units of $10^{14}~M_{\odot}$) is
scaled with $L_{500}$ and $z$ by
\begin{eqnarray}
\log M_{500} & = & 1.08\log  L_{500}-(1.50\pm0.29) \nonumber \\   
            &   & + (2.69\pm0.70)\log (1+z).
\label{m500z}
\end{eqnarray}
The richness $R_{L\ast,500}$ is thus defined as a mass proxy by
\begin{equation}
R_{L\ast,500}=L_{500}\,\Big(\frac{1+z}{1+0.7}\Big)^{2.69}, 
\label{richdef}
\end{equation}
so that the cluster mass and the richness is related by
\begin{equation}
\log M_{500}=1.08\log R_{L\ast,500}-(0.88\pm0.20).
\label{m500rich}
\end{equation}

\begin{figure*}
\resizebox{58mm}{!}{\includegraphics{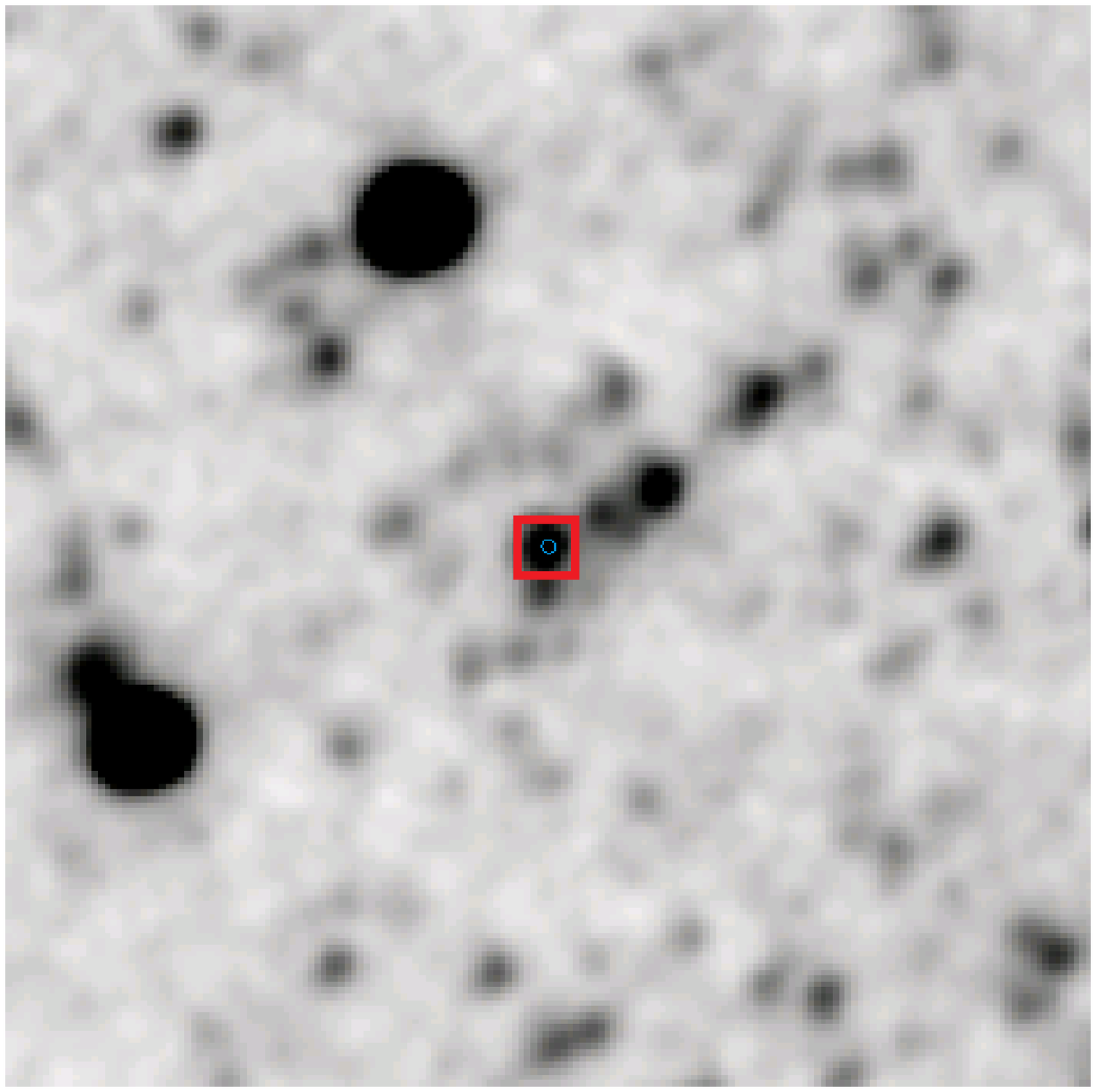}}
\resizebox{58mm}{!}{\includegraphics{f1b.eps}}
\resizebox{58mm}{!}{\includegraphics{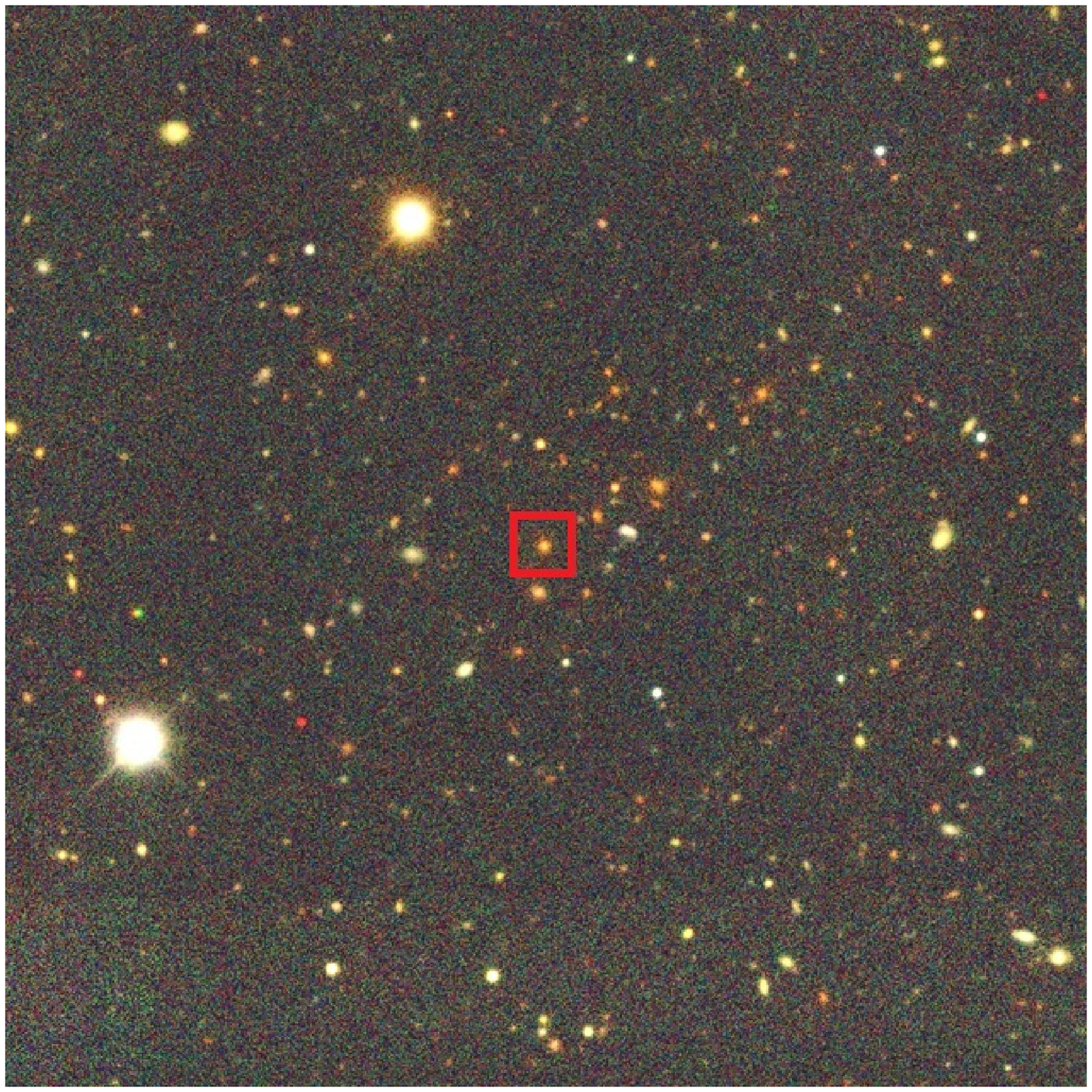}}
\caption{The $4'\times4'$ images of a rich cluster at a redshift of
  $z=0.8623$ identified from the foreground-cleaned SDSS-WISE data at
  R.A. = 4.76343$^{\circ}$ and Dec. = $-0.01448^{\circ}$. The WISE
  $W1$-band image in the left panel shows many galaxies concentrated
  around the BCG marked by the square. The black dots in the middle
  panel indicate the positions of member galaxy
  candidates, with a symbol size scaled by the square root of galaxy
  luminosities. The right panel shows the colour image of the cluster
  from the Dark Energy Survey
  (https://des.ncsa.illinois.edu/releases/dr1/dr1-access).}
\label{example}
\end{figure*}

\begin{figure}
\centering
\includegraphics[width = 0.47\textwidth]{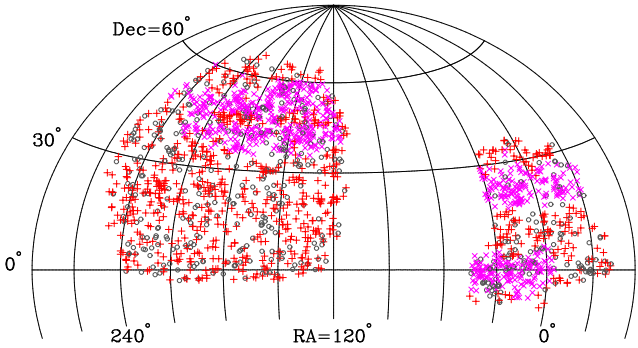}
\caption{The sky distribution of 1959 clusters of galaxies identified
  in this paper: 454 previously known clusters are indicated by
  circles, and newly identified clusters are indicated by ``+'' or
  ``$\times$'' if they are in the eBOSS region.}
\label{skycov}
\end{figure}

\begin{figure}
\centering
\includegraphics[width = 0.35\textwidth]{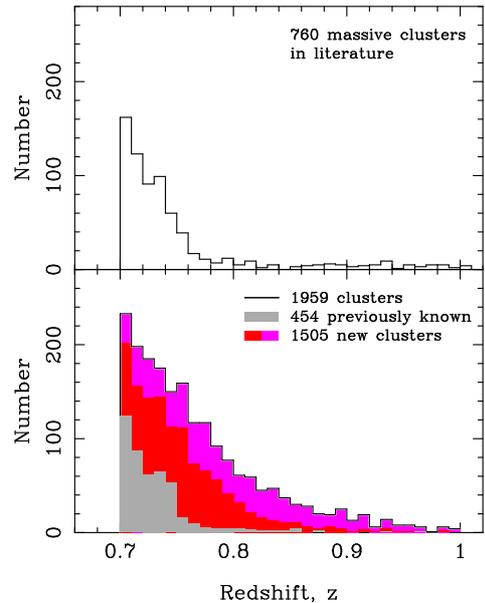}
\caption{Redshift distribution of the 1959 massive high-redshift
  clusters (lower panel), compared with all known massive clusters
  from literature (upper panel). The newly identified 1505 clusters
  significantly enlarge the high-redshift cluster sample of $z>0.75$,
  and clusters of $z>0.8$ are mostly detected in the eBOSS region
  (cf. Figure~\ref{skycov}).}
\label{hist_zc}
\end{figure}

\subsection{Identification of galaxy clusters}
\label{algorithm}

Clusters of galaxies always stand out clearly in optical/infrared data
as overdensity regions of galaxies around luminous BCGs
\citep{abe58,aco89}. Following our previous papers
\citep{whl09,wh11,wh15,why18}, we identify massive high-redshift
clusters from the cleaned SDSS-WISE data with the following steps:

1. Find an overdensity region around each BCG candidate. We take each
observed LRG as a BCG candidate, which has a spectroscopic redshift
determined already in the SDSS. We calculate the total $W1$-band
luminosity $L_{\rm 0.5}$ of member galaxy candidates within a
projected radius of 0.5~Mpc around the BCG candidate, within which a
rich cluster can have enough member galaxies \citep{gbg+95,amk+98}. We
then have to estimate the local ``background'' and also the
``fluctuation'' of the total luminosity of galaxies. For each cluster
candidate, $\langle L_{\rm 0.5}\rangle$ is estimated locally by the
galaxies within an annulus of projected distance between 2--4 Mpc from
the BCG candidate. To get the fluctuation $\sigma_{L_{\rm 0.5}}$, we
randomly select 1000 positions on the sky, and obtain the deviation of
1000 total luminosities of $L_{\rm 0.5}$. An overdensity region with a
high ``signal-to-noise ratio'', ${\rm SNR} = (L_{\rm 0.5}-\langle
L_{\rm 0.5}\rangle)/\sigma_{L_{\rm 0.5}} \ge 5$, is identified as a
cluster candidate. Obviously a larger $\rm SNR$ means a larger number
of galaxies or more luminous galaxies in the area with a photometric
redshift of $z>0.65$.

2. Estimate the cluster radius and richness of the cluster candidate
by using the scaling relations described above. We first calculate the
total $W1$-band luminosity, $L_{\rm 1Mpc}$, within a radius of 1 Mpc
around the BCG from the cleaned SDSS-WISE data, so that
$r_{500}$ can be estimated by using the
equation~(\ref{r500z}). Afterwards, the total luminosity $L_{500}$ can
be obtained from the cleaned SDSS-WISE data, and the cluster richness 
$R_{L\ast,500}$ and finally the cluster mass $M_{500}$ can be
derived by using equation~(\ref{richdef}) and equation~(\ref{m500rich}).

3. Clean the entries. It is possible that a rich cluster contains two
or more LRGs with SDSS spectroscopic redshifts, which all are regarded
as BCG candidates, so that a cluster can be sort out twice or more times in
the above procedures. We therefore perform the friends-of-friends
algorithm \citep{hg82} to merge them into one cluster if they have a
redshift difference smaller than 0.1 and a projected distance smaller
than $1.5\,r_{500}$. The BCG of the richest one is then adopted for
such a multi-identified cluster.

Figure~\ref{example} show an example of a newly identified massive
cluster, WH J001903.2$-$000052, at a redshift of $z=0.8625$ with a
richness of $R_{L\ast,500}=34.35$ and a ${\rm SNR}=9.10$.
Due to faint optical magnitudes of member galaxies, this cluster was
not identified in previous papers based on the SDSS data. The WISE
$W1$-band image is deeper for high-redshift galaxies than the SDSS
data, and shows many sources around the central BCG. After discarding
the foreground objects, the distribution of member galaxy candidates
shows a clear overdensity of galaxies around the BCG. The colour image
from Dark Energy
Survey\footnote{https://des.ncsa.illinois.edu/releases/dr1/dr1-access}
confirms that the cluster contains many member galaxies with a
similarly red colour.

In Table~\ref{tab1} we publish 1959 massive clusters identified in the
SDSS BOSS region covering $\sim$10000 square degrees (see
Figure~\ref{skycov}). The BCGs of all clusters have the spectroscopic
redshifts observed already, which are taken as the redshift of the 
clusters. The redshift distribution of the clusters is in the range of
$0.7<z<1$ as shown in Figure~\ref{hist_zc}. Comparing to the redshift
distribution of all known clusters we collected from literature in
this redshift range (including 454 known clusters re-identified in
this paper, see below), our 1505 newly identified clusters of galaxies
significantly enlarge the sample of high-redshift clusters especially
at a redshift of $z>0.75$. The identified clusters have a
richness of $R_{L\ast,500} \ge 15$ that corresponds to a mass of
$M_{500} \ge 2.5\times 10^{14}~M_{\odot}$ according to
equation~(\ref{m500rich}). Obviously, only relatively luminous cluster
members detected in the WISE $W1$-band are used for the cluster
finding. On average there are about 17 member galaxy candidates within
$r_{500}$ for each clusters (see column 12 of Table~\ref{tab1}).

\subsection{Matching with previous cluster catalogs}
\label{match_cata}

In the sky region of SDSS coverage, some of high-redshift galaxy
clusters have previously been identified from the SDSS photometric
data \citep[e.g.][]{gmb11,whl12} or deep survey data in some small
regions \citep[e.g.][]{gy05,wh11,dac+11}. We cross-match our
identified clusters of galaxies with those in previous cluster
catalogues to find out how many clusters are newly identified.

The largest catalogue that we made a few years ago comprises 132,684
clusters of $0.05<z<0.8$ from the SDSS photometric data
\citep[WHL12,][]{whl12}. By including spectroscopic redshifts of the
SDSS DR12, \citet{wh15} updated the WHL12 cluster catalogue and
identified additional 25,419 new clusters mostly with higher
redshifts, from which we get 1696 clusters with a redshift of
$z\ge0.7$.
We cross-match these 1696 clusters with 1959 clusters of this paper
within a projected separation of $1.5\,r_{500}$ and a redshift difference of
$c\Delta z/(1+z)= 2500$~km~s$^{-1}$ if spectroscopic redshifts are
available, otherwise a redshift difference of 0.05. We get
404 matches.
By adding non-overlapping clusters in other catalogues
\citep{gy05,mpe+09,wh11,dac+11,gmb11,pap+11,mrh+12,skt+12,hhm+13,
  tsl13,tsl14,bsg+15,dab+15,ltt+15,oll+18,hhs+18}, we finally have 
454 clusters previously known and the rest 1505 clusters are
newly recognized for the first time in this paper. 

As shown in Figure~\ref{hist_zc}, the identified clusters have a
median redshift of about 0.75. In the redshift range of $z>0.75$,
however, 94\% of the clusters are newly identified. Comparing with the
redshift distribution of all 760 previously known massive clusters of
$M_{500}\ge2.5\times10^{14}~M_{\odot}$ compiled from literature
\citep[references listed above, plus clusters outside the SDSS regions
  in][]{bsd+15,pcg+16,plancksz16,rrh+16}, one can see that our cluster
sample significantly enlarges the number of high-redshift
clusters. Because more LRGs with higher redshifts have been observed
in the eBOSS region, the identified clusters are about three times
more in the sky number density (about 0.28 per square degree) than
those in the other BOSS region. Most of clusters of $z>0.8$ are found
around the LRGs in the eBOSS region.

We here estimate the detection rate of clusters identified from the
cleaned SDSS-WISE data. There are 216 clusters \citep{wh15} with a
richness greater than 50 in this sky region, and the spectroscopic
redshifts of their BCGs are available. We find that 131 (61\%) of 216
clusters are detected. The detection rate is higher for richer
clusters, increasing to 71\% (76 of 107) for clusters with a WH15
richness greater than 60. Therefore, it is good to find out more
high-redshift clusters as listed Table~\ref{tab1}, while we aware that
some massive clusters are still missing because they show a lower
${\rm SNR}$ or richness in the cleaned SDSS-WISE data.

\subsection{The false detection rate}
\label{false}

Galaxy clusters in Table~\ref{tab1} have been identified from the
overdensity regions in the sky, and their member galaxies are possibly
mixed with field galaxies. It is also possible that the projection
effect from foreground or background filaments of the large scale
cosmic web may lead to a false detection of galaxy clusters.
Therefore, the false detection rate should be estimated for a cluster
sample identified from optical/infrared photometric data.

\begin{figure}
\centering \includegraphics[width = 0.4\textwidth]{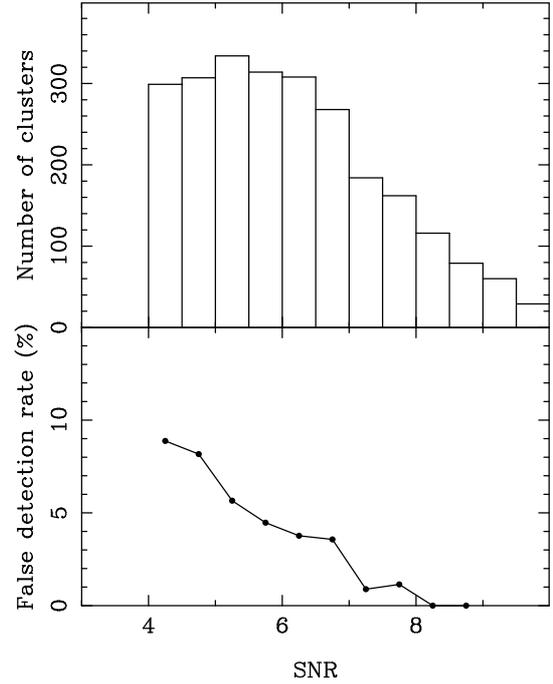}
\caption{Distribution of cluster detection $\rm SNR$ (upper panel) and 
false detection rate as a function of $\rm SNR$ (lower panel).}
\label{falserate}
\end{figure}

We estimate the false detection rate by using a mock galaxy catalogue
generated from the observed data
\citep[e.g.,][]{gsn+02,kma+07b,hmk+10,whl12}. We start from the
catalogue of WISE galaxies, in which low-redshift galaxies are kept
temporally. We first remove the member galaxy candidates of clusters
in this work and those of clusters in \citet{wh15} from the galaxy
catalogue. Then, we generate a mock catalogue by randomly shuffling
the magnitudes and photometric redshifts together for the rest
galaxies, so that the positions of galaxies and the correlation
between redshift and magnitude are all reserved. Subsequently we
remove the low-redshift galaxies of $z<0.65$. These procedures can
eliminate real unidentified clusters, but the two dimensional
distribution of galaxies are kept the same as the real sample.
Afterwards, the BCG candidates with spectroscopic redshifts are
randomly put on the sky, so that we can check if the projection effect
leads to any false detections. We apply the same cluster
identification procedures discussed above to the mock data, with the
thresholds of ${\rm SNR} \ge 4$ and $R_{L*,500} \ge 15$. Any detected
``clusters'' around the BCG candidates from the mock data therefore
can be regarded as false detections. The false detection rate is
calculated as being the ratio between the number of false detections
from the mock data and the total number of identified clusters from
the real data. Here, the false detection rate with $4\le{\rm SNR}<5$
is also calculated. To minimize the random noise, we generate ten such
mock samples of galaxies and get an average false detection rate.

As shown in Figure~\ref{falserate}, the false detection rate depends
on the signal-to-noise ratio $\rm SNR$ of the overdensity
detection. It is negligible for $\rm SNR>8$, but increases with a
smaller $\rm SNR$. We therefore set the threshold of ${\rm SNR}\ge5$
so that the false detection rate is $\lesssim 5\%$ in each bin, and
about 3\% for the whole sample.

\section{Properties of clusters}

Combining the clusters we identified at high redshifts in this paper
with the massive clusters we previously identified at lower
redshifts from the SDSS data, we can investigate the evolution of BCGs
and cluster properties.

\begin{figure}
\centering
\includegraphics[width = 0.4\textwidth]{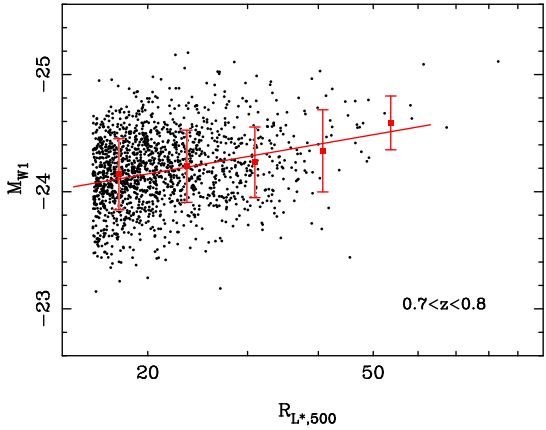}
\caption{The correlation between the $W1$-band absolute magnitude of
  BCGs and cluster richness confirms that richer clusters host more
  luminous BCGs.}
\label{bcgrich}
\end{figure}

\subsection{Evolution of BCGs}

BCGs are the most massive galaxies located at the potential centres of
clusters. Generally, they have an elliptical morphology but show
different properties from normal elliptical galaxies
\citep[][]{sch86,bhs+07,lxm+08}. Optical and near-infrared data show
that the color evolution of BCGs is in good agreement with a passive
population synthesis model \citep{ses+08,wad+08,wh11}. However, several
studies show signatures of ongoing star formation or active galactic
nuclei (AGN) in some BCGs
\citep{cae+99,mrb+06,lmm12,fbp+14,ges+16,dcv+17,bwm+17}.

Previously, we found that richer clusters host more luminous BCGs at
redshifts $z<0.42$ \citep{whl12}. Here we can investigate if it keeps
true at a much higher redshift. For the clusters of $0.7<z<0.8$, the
$W1$-band absolute magnitudes of BCGs are obtained and then plotted 
against the cluster richness as shown in Figure~\ref{bcgrich}. 
We find the best fit to the data as being 
\begin{equation}
M_{W1}=(-23.02\pm0.24)-(0.87\pm0.16)\log R_{L\ast,500}, 
\end{equation}
which is very consistent with those for low-redshift clusters shown 
in figure 19 of \citet{whl12}.

\begin{figure}
\centering
\includegraphics[width = 0.4\textwidth]{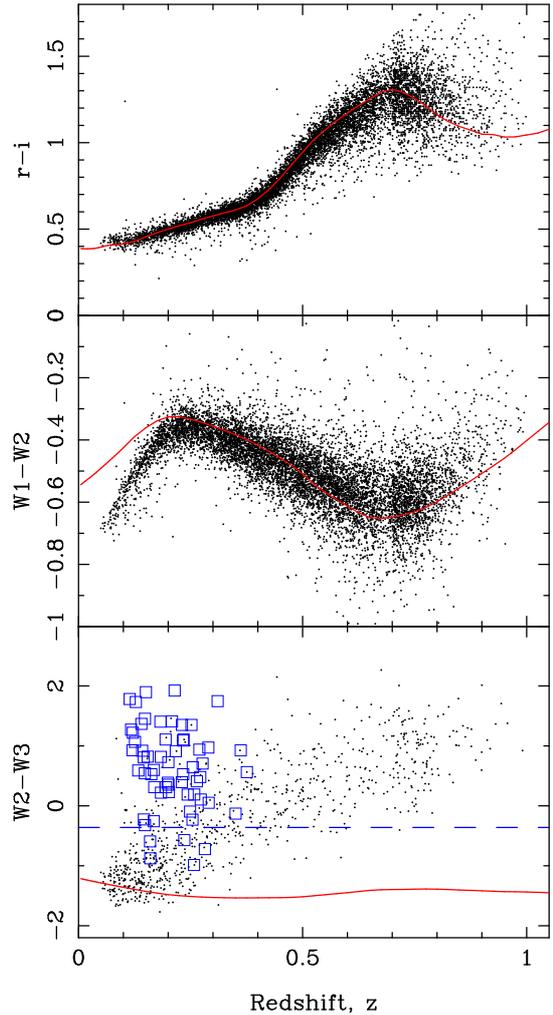}
\caption{Evolution of colours of $r-i$ (top panel), $W1-W2$
  (middle panel) and $W2-W3$ (bottom panel) as a function of redshift
  for 9192 BCGs in massive clusters. The solid lines represent the
  colour evolution calculated by using BC03 model in which stellar
  population was formed at $z_f=3$ with passive stellar evolution
  tracks. In the bottom panel, only 835 BCGs are shown, and others are
  below the detection limit in $W3$ band. Above the dashed line are
  galaxies probably with ongoing star formation, which is verified by
  59 BCGs from \citet{lmm12} indicated by squares.}
\label{colorz}
\end{figure}

\citet{wh11} studied the optical colour evolution of BCGs by using a
sample of clusters up to a redshift of $z\sim1.6$. Here, we
investigate the optical and mid-infrated colour evolution of BCGs in
the redshift range of $0.05<z<1$ by using a combined sample of 9192
BCGs of massive clusters from this work and \citet[][with a richness
  $\ge 50$]{wh15}.  We get the magnitude data of all these BCGs in the
$r$, $i$, $W1$ and $W2$ bands, which should be dominated by old
stellar populations. The magnitude data in the $W3$ band, however, is
sensitive to star formation or nuclei activity \citep{ydt+13,jmt+13},
and they are available for only 835 of the 9192 BCGs with a detection
threshold of signal-to-noise ratio greater than 3.

We first show the redshift evolution of the optical colour $r-i$ and
mid-infrared colour $W1-W2$ for the BCGs in Figure~\ref{colorz}.  To
explain colour evolution, we calculate the colours by using a passive
population synthesis model \citep[][hereafter BC03]{bc03}, in which
stellar population was formed at a redshift of $z_f=3$. We adopt the
stellar evolution tracks of Padova 1994 \citep{gbc+96}, the `Basel3.1'
stellar spectral library \citep{wlb+02}, and the initial mass function
of \citet{cha03} and the Solar metallicity in the model. We find that
the colour data $r-i$ are in good agreement with the BC03 model. The
distribution of mid-infrared colour $W1-W2$ shows a peak at the
redshift of $z\sim0.2$ and a valley at the redshift of
$z\sim0.7$. The colour data $W1-W2$ are marginally consistent with the BC03
model at redshifts $z>0.2$, but are significantly lower than the model
at redshifts $z\le0.2$. We noticed that the BC03 model was calibrated
by using only the observed optical and near-infrared data
\citep{bc03}.  Obviously the mid-infrared WISE data here can provide
good constraints on the population synthesis model of galaxies, which
is beyond the scope of this paper.

In the bottom panel of Figure~\ref{colorz}, we show the colour
evolution of $W2-W3$ for the 835 BCGs, which are marginally consistent
with the BC03 model only at redshifts $z\le0.2$ but not at higher
redshifts. For reference, we here take 59 BCGs from \citet {lmm12},
detected in the $W3$ band showing ongoing star formation with a star
formation rate greater than 1 $M_{\odot}$/yr. We find that 54 of the
59 BCGs have a colour of $W2-W3>-0.36$, which we adopt here as a
criteria for deviating from the passive model \citep[e.g.][]{ges+16}.
Figure~\ref{colorz} shows that 428 of the 835 BCGs have a colour of
$W2-W3>-0.36$. The rest BCGs without $W3$ detection may or may not
have violent activities but can not be judged probably due to their
high redshifts. We therefore conclude that at least 428 of 9192 ($\sim
4.6\%$) of BCGs in massive clusters have ongoing star formation or
active nuclei.

\subsection{Evolution of the fraction of blue galaxies}

By using the clusters of $0.1<z<0.4$ from the SDSS \citep{wh15} and
the high-redshift clusters in this work, we verify the Butcher--Oemler
effect, i.e. clusters at higher redshifts contain more blue member 
galaxies than those at lower redshifts 
\citep[e.g.][]{bo78,bo84,goy+03,hse+09,lsb+11}.

\begin{figure}
\centering
\includegraphics[width = 0.4\textwidth]{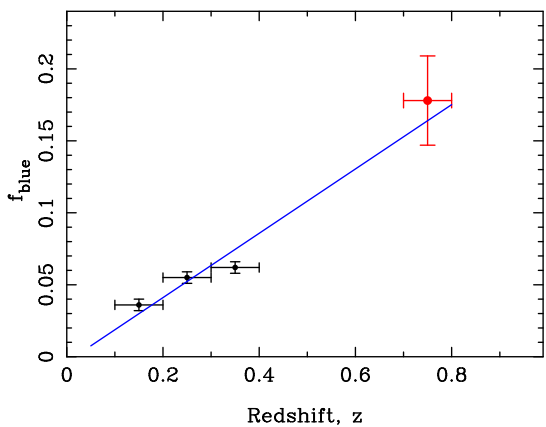}
\caption{The fraction of blue member galaxies in clusters evolves as a
  function of redshift. The values at low redsdhift ($z<0.4$, black
  dots) are obtained by using the SDSS data, and that at $z=0.75$ by 
  using the CFHT data. The solid line is the best fit to the data.}
\label{fracb}
\end{figure}

Considering that cluster galaxies of $M^e_r\le -20.5$ are
volume-limited complete to a redshift of $z\sim0.4$ in the SDSS data,
here $M^e_r$ is the $r$-band evolution-corrected absolute magnitude
defined in \citet{whl12}, we take data of member galaxies of 2943
massive clusters with a richness $\ge50$ in the redshift range of
$0.1<z<0.4$ \citep{wh15}. Those clusters have a comparable mass with
the clusters we identified in this paper.
We use the photometric redshifts to discriminate member galaxies of
$M^e_r\le -20.5$ (spectroscopic redshifts are used if available).  The
photometric redshifts have been estimated for the SDSS galaxies to a
limit of $r=21.5$ with an uncertainty of $\sigma_z \simeq 0.023(1+z)$
\citep{bcl+14}. Member galaxies are obtained if they have a
photometric redshift within a redshift slice of 1.6\,$\sigma_z$ from a
cluster redshift, so that about 90\% member galaxies can be included
within the redshift slice. Blue galaxies is defined to those with a
rest-frame colour of $g-r$ bluer than the colour of cluster red
sequence by 0.2 mag \citep{goy+03}. We get the number of blue galaxies
and the number of all galaxies in a cluster, after subtraction of the
backgrounds estimated by using galaxies within the same redshift slice
and the annulus of 2--4 Mpc from a cluster centre. We stack clusters
in three redshift bins to get the averaged fractions of blue galaxies,
as shown in Figure~\ref{fracb}.

For the high-redshift clusters identified in this paper, we use the
deeper CFHT Legacy Survey
data\footnote{http://www.cfht.hawaii.edu/Science/CFHLS/} to get their
member galaxies. The latest T0007 data
release\footnote{http://terapix.iap.fr/} covers about 155 square
degrees in four independent fields \citep{hcw+12}, reaching a limit of
$i\approx23.5$ for extended sources (80\% completeness) that is about
2 magnitude deeper than the SDSS data. The absolute magnitudes are
available, and the galaxy photometric redshifts were estimated based
on the methods of \citet{iam+06} and \citet{cik+09} with a redshift
uncertainty of $0.033(1+z)$. The four CFHT fields are overlapped with
the SDSS survey and we get member galaxies for 29 clusters of $0.7<z<0.8$
from the CFHT data. Following \citet{goy+03}, we find a fraction of
blue galaxies $f_{\rm blue}=0.18\pm0.03$ at $z\sim0.75$ (see
Figure~\ref{fracb}), much higher than those in clusters of $z<0.4$,
indicating that the blue fraction increases with redshift to $z\sim0.8$.
It is interesting to see $f_{\rm blue} = (0.22\pm0.03)z\pm0.01$ for
those massive clusters with the blue galaxies criterion setting as
bluer $g-r$ than the red sequence by 0.2 mag. However, we notice that
in addition to the redshift evolution effect, the blue fraction
depends on cluster mass, cluster-centric distance and galaxy magnitude
\citep{pfc+05,hsw+09}, which may explain the discrepancy between our
result and the larger fractions obtained at similar redshifts in some
previous works \citep[e.g.][]{goy+03,dpa+07,noo+18}.

\begin{figure}
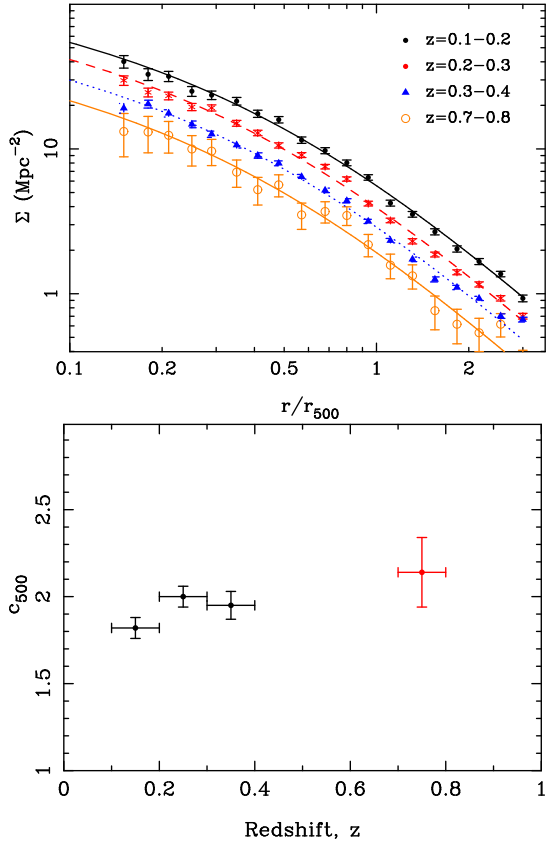

\centering
\includegraphics[width = 0.4\textwidth]{f8a.eps}
\includegraphics[width = 0.4\textwidth]{f8b.eps}
\caption{Surface number density profiles of member galaxies in
  different redshift bins are fitted with a projected NFW profile
  (upper panel). To show clearly, data points and fittings are shifted
  in the vertical direction by multiplying a factor of $10^{-0.1}$,
  $10^{-0.2}$ and $10^{-0.3}$ for $z=0.2$--0.3, $z=$0.3--0.4 and
  $z=0.7$--0.8, respectively. The concentration parameter is plotted
  against redshift in the lower panel, showing no significant
  evolution.}
\label{concen}
\end{figure}

\subsection{Surface number density profile of member galaxies}

The member galaxies of 2943 massive clusters with a richness $\ge50$
in the redshift range of $0.1<z<0.4$ and of 29 clusters of $0.7<z<0.8$
from the CFHT data can be used to investigate the evolution of surface
number density profile of member galaxies. The three-dimensional
number density profile of member galaxies can be described by the NFW
model \citep{nfw+97}
\begin{equation}
\rho(r)=\frac{\rho_s}{\Big(r/r_s\Big)\Big(1+r/r_s\Big)^2},
\end{equation}
where $\rho_s$ and $r_s$ are the characteristic density and radius,
respectively. The concentration parameter for the NFW profile is
defined as $c_{500}=r_{500}/r_s$. The projected surface density
profile, $\Sigma_{\rm NFW}$, is then obtained by integrating the
profile along the light of sight \citep{bar96}.

We stack the clusters in four redshift bins and calculate the surface
number density profiles of member galaxies out to the distance of
$3\,r_{500}$, which are then fitted with the superposition of a
projected NFW profile and a constant background
\begin{equation}
\Sigma(r/r_{500})=\Sigma_{\rm  NFW}+\Sigma_{\rm  back}, 
\end{equation}
as shown in Figure~\ref{concen}. We find that the number density
profile has no significant redshift evolution, consistent with the
recent conclusions based on the SPT-SZ clusters of $z<1.1$
\citep{zmd+16,hmz+17} and the clusters of $z<1$ in the Subaru HSC
Survey \citep{lhl+17,noo+18}. The concentration parameter of
$c_{500}\sim2$ is in agreement with those found by \citet{hmz+17} and
\citet{lhl+17}.

\section{Summary}

We identify a sample of 1959 massive galaxy clusters in the redshift
range of $0.7<z<1.0$ by using the data from the SDSS and WISE, among
which 1505 clusters are recognized for the first time. These clusters
are identified around SDSS LRGs as being the overdensity regions of
galaxies with a signal-to-noise ratio greater than 5. This cluster
sample significantly enlarges the number of massive clusters at
$z>0.75$.
We combine these high-redshift clusters with the clusters in a lower
redshift range to study the evolution of BCGs and clusters. The data
verify the Butcher--Omler effect, and show that richer clusters have
more luminous BCGs even at a high redshift. A small fraction of BCGs
show activities of star formation or active nuclei. The profile of
number density of member galaxies in these massive clusters does not
evolve with redshift.

\section*{Acknowledgements}

We thank the referee for valuable comments that helped to improve the
paper. The authors are partially supported by the National Natural Science
Foundation of China (Grant No. 11473034, U1731127), the Key Research
Program of the Chinese Academy of Sciences (Grant
No. QYZDJ-SSW-SLH021) and the strategic Priority Research Program of
Chinese Academy of Sciences (Grant No. XDB23010200), and the Open
Project Program of the Key Laboratory of FAST, NAOC, Chinese Academy
of Sciences.
Funding for the Sloan Digital Sky Survey IV has been provided by the
Alfred P. Sloan Foundation, the U.S. Department of Energy Office of
Science, and the Participating Institutions. SDSS acknowledges support
and resources from the Center for High-Performance Computing at the
University of Utah. The SDSS web site is www.sdss.org.
This publication makes use of data products from the Wide-field
Infrared Survey Explorer, which is a joint project of the University
of California, Los Angeles, and the Jet Propulsion
Laboratory/California Institute of Technology, funded by the National
Aeronautics and Space Administration.
This work is based in part on data products produced at Terapix
available at the Canadian Astronomy Data Centre as part of the
Canada-France-Hawaii Telescope Legacy Survey, a collaborative project
of NRC and CNRS, and also based on observations obtained with
MegaPrime/MegaCam, a joint project of CFHT and CEA/IRFU, at the
Canada-France-Hawaii Telescope (CFHT) which is operated by the
National Research Council (NRC) of Canada, the Institut National des
Science de l'Univers of the Centre National de la Recherche
Scientifique (CNRS) of France, and the University of Hawaii.

\bibliographystyle{mnras}
\bibliography{wise}

\begin{appendix}

\section{Scaling relations obtained from 45 known clusters}
\label{app}

\begin{figure*}
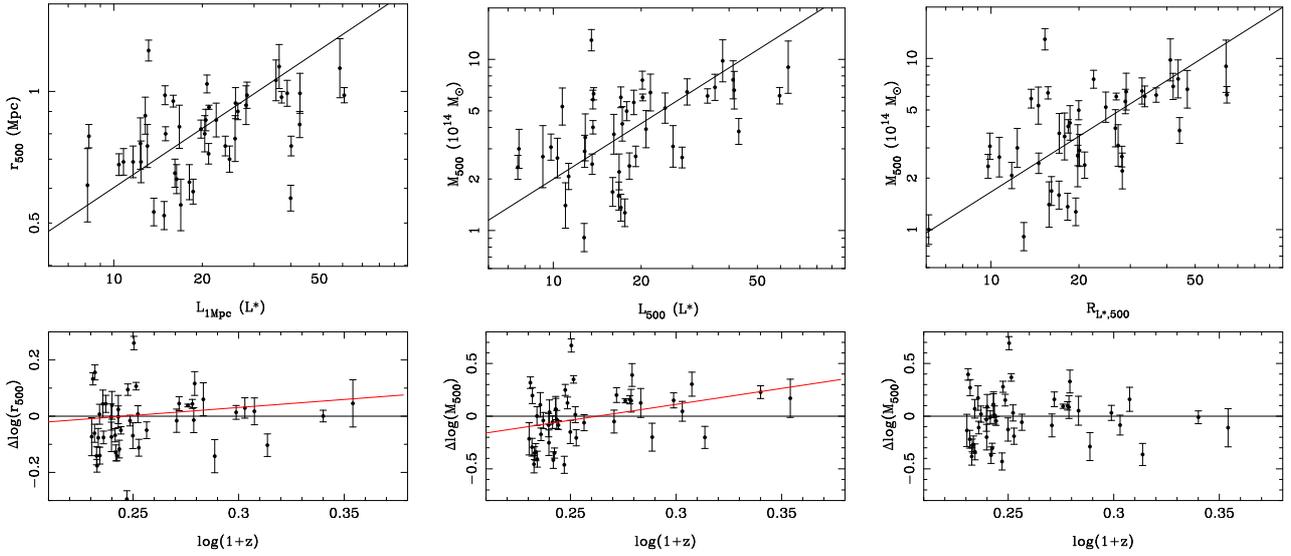

\centering
\includegraphics[width=0.3\textwidth]{fa1a.eps} \hspace{3mm}
\includegraphics[width=0.3\textwidth]{fa1b.eps} \hspace{3mm}
\includegraphics[width=0.3\textwidth]{fa1c.eps}\\[1mm]
\includegraphics[width=0.3\textwidth]{fa1d.eps} \hspace{3mm}
\includegraphics[width=0.3\textwidth]{fa1e.eps} \hspace{3mm}
\includegraphics[width=0.3\textwidth]{fa1f.eps}
\caption{Upper panels show the scaling relations between cluster
  parameters derived from the WISE data (the X-axis) and cluster
  radius or mass (the Y-axis) for the 45 compiled
  clusters. The data deviations from the scaling relation are
  plotted in the lower panels to check the redshift dependence.}
\label{scaleXO}
\end{figure*}

To get the scaling relations for cluster radius and mass from
observational data, we compile the cluster masses $M_{500}$ derived by
X-ray or SZ measurements for 45 known clusters from literature
\citep{lhk+08,heb+08,vbe+09,mae+10,rbf+11,pap+11,
  hhm+13,tsl13,tsl14,gad+14, plancksz13,gdb+15,bgl+15,bsg+15,
  plancksz16,tdm+16}, which are located in the SDSS region within a
redshift range of $0.7\le z\lesssim 1$. The average mass is taken if a
cluster has two or more mass estimates available in literature.

The location of a BCG is taken as the cluster centre, and its redshift
as the redshift of a cluster. For each galaxy cluster, we recognize 
the member galaxy candidates from the cleaned SDSS-WISE data to 
calculate a total $W1$-band luminosity. Galaxies with a
spectroscopic redshift within a redshift slice of $c\Delta z/(1+z)=
2500$~km~s$^{-1}$ from the cluster redshift, $z$, are considered as
member galaxies directly. Here $c$ is the speed of light. Other
galaxies in the cleaned WISE catalogue have no redshifts, and are
adopted as cluster member galaxy candidates only if they are fainter
than the BCG, because all stars and galaxies with a lower
photometric redshift have been cleaned already. The total $W1$-band
luminosity is then easily calculated by summing the luminosities of
all member galaxy candidates. Follow the method in our previous papers
\citep{whl12,wh15}, the local background is estimated from an annulus
of projected distance between 2--4 Mpc from the BCG candidate, and
then subtracted from the total $W1$-band luminosity.

We calculate the total $W1$-band luminosity within a radius of 1 Mpc
from the BCG, $L_{\rm 1Mpc}$, in units of $L^{\ast}$ which
is the evolved characteristic luminosity of galaxies in the $W1$ band
\citep{mgb+10}. The left panel of Figure~\ref{scaleXO} shows the
correlation between cluster radius $r_{500}$ and $L_{\rm 1Mpc}$ for
the 45 compiled clusters. Adopting a slope of 0.45 that was obtained
by \citet{wh15}, we find that the best fit to the data is
\begin{equation}
\log r_{500}=0.45\log L_{\rm 1Mpc}-(0.67\pm0.10),
\label{r500}
\end{equation}
where $r_{500}$ is in units of Mpc and $L_{\rm 1Mpc}$ is in units of
$L^{\ast}$.
Noted that we use the flux-limited sample of WISE galaxies to
calculate the total luminosity. Clusters at lower redshifts should
contain more member galaxies than those at higher redshifts. The total
luminosities of member galaxy candidates are then biased to smaller
values at higher redshifts, so that the above relation may have a
redshift dependence to check. We get the deviations of $r_{500}$ from
the above fitted relation, $\Delta\log r_{500}=\log
r_{500}-(0.45\,\log L_{\rm 1Mpc}-0.67)$, and plot them against
redshift. We find a slight dependence on redshift in the form of
\begin{equation}
\Delta\log r_{500}=(0.57\pm0.29)\log (1+z)-(0.14\pm0.07).
\end{equation}
Considering this correction, the radius $r_{500}$ is related to
$L_{\rm 1Mpc}$ and $z$ by
\begin{eqnarray}
\log r_{500}&=&0.45\log L_{\rm 1Mpc}-(0.81\pm0.12)\nonumber\\
&&+(0.57\pm0.29)\log (1+z).
\end{eqnarray}
Similar, we calculate the total $W1$-band luminosity $L_{500}$ (also
in units of $L^{\ast}$) within $r_{500}$ for the 45 compiled clusters,
certainly with a background subtraction. The cluster mass $M_{500}$ and
$L_{500}$ are found to be closely correlated, as shown in the
middle panel of Figure~\ref{scaleXO}. Adopting a slope of 1.08 that
was obtained by \citet{wh15}, we find that the best fit to the data is
\begin{equation}
\log M_{500}=1.08\log  L_{500}-(0.78\pm0.23),
\end{equation}
where $M_{500}$ is in units of $10^{14}~M_{\odot}$. Again, to correct
the redshift dependence, we get the deviations of
$M_{500}$ from the above fitted relation, $\Delta\log M_{500}=\log
M_{500}-(1.08\,\log L_{500}-0.78)$, and plot them against redshift.
We find a significant dependence on redshift in the form of
\begin{equation}
\Delta\log M_{500}=(2.69\pm0.70)\log (1+z)-(0.72\pm0.17).
\end{equation}
Combining the last two equations, we get
\begin{eqnarray}
\log M_{500} & = & 1.08\log  L_{500}-(1.50\pm0.29) \\   
            &   & + (2.69\pm0.70)\log (1+z).
\end{eqnarray}

We define the cluster richness by using the total $W1$-band luminosity
$L_{500}$ as being
\begin{equation}
  R_{L\ast,500}=L_{500}\,\Big(\frac{1+z}{1+0.7}\Big)^{2.69}.
\end{equation}
The scaling relation between the cluster mass and the richness is
given by
\begin{equation}
\log M_{500}=1.08\log R_{L\ast,500}-(0.88\pm0.20), 
\end{equation}
which seems have no redshift bias as shown in the right panels of 
Figure~\ref{scaleXO}.

\begin{table*}
\begin{minipage}{160mm}
\begin{center}
\caption[]{Parameters of 45 clusters with known mass ($M_{\rm 500}$) from literature}
\begin{tabular}{rrrrcccrr}
\hline
\mc{1}{l}{No.}    &\mc{1}{c}{RA}       &\mc{1}{c}{Dec.}   &\mc{1}{c}{$z$}      & \mc{1}{c}{$M_{\rm 500}$} &\mc{1}{c}{$r_{500}$}& 
\mc{1}{c}{Ref.}&\mc{1}{c}{$L_{\rm 1Mpc}$}&\mc{1}{c}{$L_{500}$} \\
(1) & \mc{1}{c}{(2)} & \mc{1}{c}{(3)} & \mc{1}{c}{(4)} & (5) & \mc{1}{c}{(6)} & \mc{1}{c}{(7)} & (8) & \mc{1}{c}{(9)} \\
\hline
   1 &   3.05893 &  16.03893 & 0.9440 & $ 1.40\pm0.50$ & $0.55\pm0.08$ & B15a    & 16.90 & 11.01  \\
   2 &   5.55431 &  -0.60944 & 0.8050 & $ 6.86\pm1.32$ & $0.99\pm0.07$ & H13     & 38.89 & 35.78  \\
   3 &  14.78542 &  -0.83494 & 0.7877 & $ 6.45\pm1.24$ & $0.98\pm0.07$ & H13     & 28.39 & 28.73  \\
   4 &  19.99224 &   0.92608 & 0.7381 & $ 3.65\pm1.12$ & $0.83\pm0.10$ & H13     & 16.71 & 16.13  \\
   5 &  31.55611 &  -1.23810 & 0.7146 & $ 5.19\pm1.17$ & $0.94\pm0.08$ & H13     & 25.94 & 24.15  \\
   6 &  33.86876 &   0.51044 & 0.8650 & $ 3.91\pm1.11$ & $0.80\pm0.08$ & H13     & 20.38 & 20.78  \\
   7 &  36.14264 &  -0.04238 & 0.7725 & $ 4.98\pm0.68$ & $0.90\pm0.04$ & H08     & 26.42 & 17.84  \\
   8 &  37.12687 &   0.50991 & 0.7208 & $ 2.65\pm0.81$ & $0.75\pm0.09$ & H13,T16 & 13.01 & 10.33  \\
   9 & 127.61035 &  52.69312 & 0.9900 & $ 5.60\pm1.00$ & $0.86\pm0.05$ & L08     & 20.56 & 18.87  \\
  10 & 132.24341 &  44.86574 & 1.2600 & $ 2.70\pm1.40$ & $0.61\pm0.13$ & R11     &  8.14 &  9.21  \\
  11 & 137.86015 &   5.83725 & 0.7682 & $ 7.56\pm0.97$ & $1.04\pm0.05$ & P15     & 20.74 & 20.19  \\
  12 & 147.04832 &  29.11922 & 0.7780 & $ 6.60\pm1.90$ & $0.99\pm0.11$ & B15b    & 42.96 & 41.66  \\
  13 & 155.64725 &  13.18128 & 0.7053 & $ 1.68\pm0.36$ & $0.65\pm0.05$ & T14     & 16.14 & 15.97  \\
  14 & 156.95279 &   0.06054 & 0.7060 & $ 3.07\pm0.59$ & $0.79\pm0.05$ & T14     &  8.23 &  9.82  \\
  15 & 162.01755 &  31.64720 & 0.7500 & $ 9.80\pm3.20$ & $1.14\pm0.14$ & B15b    & 36.52 & 38.04  \\
  16 & 163.15559 &  57.51781 & 0.7090 & $ 0.91\pm0.19$ & $0.52\pm0.04$ & T14     & 14.83 & 12.76  \\
  17 & 175.09245 &  66.13752 & 0.7843 & $ 6.29\pm0.54$ & $0.95\pm0.03$ & P11,G14 & 15.93 & 13.75  \\
  18 & 175.69792 &  15.45319 & 1.1880 & $ 6.00\pm0.90$ & $0.82\pm0.04$ & G15     & 19.79 & 17.04  \\
  19 & 178.94098 &  39.02004 & 1.0090 & $ 2.90\pm0.70$ & $0.69\pm0.06$ & B15a    & 11.63 & 12.81  \\
  20 & 182.49605 &  49.89782 & 0.9020 & $ 5.30\pm1.50$ & $0.88\pm0.09$ & B15b    & 12.80 & 10.74  \\
  21 & 186.68127 &  33.76685 & 0.7665 & $ 1.27\pm0.26$ & $0.57\pm0.04$ & T14     & 39.95 & 17.59  \\
  22 & 186.74268 &  33.54683 & 0.8880 & $ 5.99\pm0.28$ & $0.92\pm0.01$ & V09,M10 & 21.07 & 20.24  \\
  23 & 187.69325 &  10.94824 & 0.7519 & $ 1.00\pm0.22$ & $0.53\pm0.04$ & T14     & 13.68 &  5.64  \\
  24 & 187.71085 &  41.57208 & 0.7454 & $ 1.36\pm0.27$ & $0.59\pm0.04$ & T14     & 18.61 & 17.04  \\
  25 & 190.80023 &  13.21965 & 0.7896 & $ 2.38\pm0.46$ & $0.70\pm0.05$ & T14     & 24.76 & 18.22  \\
  26 & 198.41559 &  22.19755 & 0.7370 & $ 3.10\pm1.00$ & $0.78\pm0.10$ & B15b    & 25.86 & 25.71  \\
  27 & 201.20351 &  30.19414 & 0.7550 & $ 6.11\pm0.62$ & $0.97\pm0.03$ & G14     & 37.21 & 33.75  \\
  28 & 204.38356 &  19.97465 & 0.9000 & $ 6.40\pm1.80$ & $0.93\pm0.10$ & B15b    & 28.17 & 21.52  \\
  29 & 204.71948 &   4.72021 & 0.7264 & $ 2.07\pm0.40$ & $0.69\pm0.05$ & T14     & 10.79 & 11.29  \\
  30 & 205.70715 &  40.47372 & 0.7099 & $ 2.67\pm0.40$ & $0.75\pm0.04$ & P11     & 40.11 & 27.62  \\
  31 & 205.76897 &  -0.01547 & 0.7151 & $ 3.79\pm0.72$ & $0.84\pm0.06$ & T14     & 42.87 & 43.17  \\
  32 & 208.43817 &  43.48428 & 0.7365 & $ 7.59\pm2.21$ & $1.06\pm0.12$ & B15b,P16& 35.62 & 41.28  \\
  33 & 213.79633 &  36.20111 & 1.0300 & $ 3.00\pm0.90$ & $0.69\pm0.08$ & R11     & 12.37 &  7.63  \\
  34 & 217.27792 &  42.68572 & 0.9200 & $ 3.50\pm1.30$ & $0.76\pm0.11$ & R11     & 12.32 & 12.87  \\
  35 & 217.80191 &  -0.10467 & 0.7122 & $ 1.59\pm0.33$ & $0.63\pm0.05$ & T14     & 16.35 & 16.76  \\
  36 & 224.76913 &  52.81554 & 0.7023 & $ 5.82\pm0.78$ & $0.98\pm0.05$ & P16     & 14.93 & 13.67  \\
  37 & 228.67696 &  13.78289 & 1.0590 & $ 2.20\pm0.60$ & $0.62\pm0.06$ & B15a    & 18.07 & 16.81  \\
  38 & 229.48769 &  31.45816 & 0.7440 & $ 2.44\pm0.36$ & $0.72\pm0.04$ & P11     & 21.01 & 13.59  \\
  39 & 231.63821 &  54.15210 & 0.7476 & $ 6.13\pm0.72$ & $0.98\pm0.04$ & P16     & 60.76 & 59.59  \\
  40 & 231.92551 &  20.74407 & 0.7000 & $ 9.00\pm3.80$ & $1.13\pm0.19$ & B15b    & 58.73 & 63.78  \\
  41 & 234.39471 &  38.48075 & 0.7500 & $ 4.20\pm1.10$ & $0.86\pm0.08$ & B15b    & 22.30 & 17.23  \\
  42 & 245.04208 &  29.49007 & 0.8700 & $ 2.34\pm0.41$ & $0.68\pm0.04$ & H08     & 10.41 &  7.56  \\
  43 & 345.70044 &   8.73083 & 0.7220 & $ 2.71\pm0.40$ & $0.75\pm0.04$ & P11     & 23.95 & 19.15  \\
  44 & 349.63193 &   0.57303 & 0.7800 & $12.94\pm2.00$ & $1.24\pm0.07$ & H08     & 13.13 & 13.51  \\
  45 & 349.97263 &   0.63699 & 0.8972 & $ 4.01\pm0.38$ & $0.80\pm0.03$ & H08     & 15.00 & 13.68  \\
\hline
\end{tabular}
\end{center}
{Note. 
Column (1): cluster number sequence;
Column (2)--(4): Right ascension, Declination (J2000) and redshift of a cluster;
Column (5): cluster mass, $M_{500}$ in units of $10^{14}~M_{\odot}$;
Column (6): cluster radius, $r_{500}$ in Mpc;
Column (7): References for cluster mass: L08 for \citet{lhk+08}, H08 for \citet{heb+08}, V09 for \citet{vbe+09},
M10 for \citet{mae+10}, P11 for \citet{pap+11}, R11 for \citet{rbf+11}, H13 for \citet{hhm+13}, T14 for \citet{tsl13,tsl14}, G14 for \citet{gad+14}, P15 \citet{plancksz13}, G15 for \citet{gdb+15},
B15a for \citet{bgl+15}, B15b for \citet{bsg+15}, P16 for \citet{plancksz16} and T16 for \citet{tdm+16};
Column (8): $W1$-band total luminosity within a radius of 1 Mpc in units of $L^{\ast}$;
Column (9): $W1$-band total luminosity within $r_{500}$ in units of $L^{\ast}$.}
\label{tab_app}
\end{minipage}
\end{table*} 

\end{appendix}

\label{lastpage}
\end{document}